\documentclass[11pt, conference]{IEEEtran}
\makeatletter
\def\blfootnote{\xdef\@thefnmark{}\@footnotetext}
\makeatother

\usepackage{cite}
\usepackage{url}

\usepackage{graphicx}
\usepackage{algorithm}
\usepackage[noend]{algorithmic}
\usepackage{subfig}
\usepackage{amssymb, amsmath,graphicx,charter, latexsym}
\usepackage{enumerate}

\newtheorem{definition}{Definition}

\newtheorem{theorem}{Theorem}
\newtheorem{example}{Example}

\def\baselinestretch{0.83}
\begin{document}
\title{A Non-Monetary Protocol for Peer-to-Peer Content Distribution in Wireless Broadcast Networks with Network Coding}
\author{
\IEEEauthorblockN{I-Hong Hou, Yao Liu, and Alex Sprintson}
\IEEEauthorblockA{CESG and Department of ECE\\Texas A\&M University\\College Station, TX 77843, USA\\Email:\{ihou,steveliu,spalex\}@tamu.edu}
%\and \IEEEauthorblockN{Yao Liu}
%\IEEEauthorblockA{CESG and Department of ECE\\Texas A\&M University\\College Station, TX 77843, %USA\\steveliu@tamu.edu}
%\and \IEEEauthorblockN{}
%\IEEEauthorblockA{CESG and Department of ECE\\Texas A\&M University\\College Station, TX 77843, %USA\\spalex@tamu.edu}
}
\maketitle\blfootnote{}

\begin{abstract}
 This paper studies the problem of content distribution in wireless peer-to-peer networks where all nodes are selfish and non-cooperative. We propose a model that considers both the broadcast nature of wireless channels and the incentives of nodes, where each node aims to increase its own download rate and reduce its upload rate through the course of content distribution. We then propose a protocol for these selfish nodes to exchange contents. Our protocol is distributed and does not require the exchange of money, reputation, etc., and hence can be easily implemented without additional infrastructure. Moreover, we show that our protocol can be easily modified to employ network coding.

 The performance of our protocol is studied. We derive a closed-form expression of Nash Equilibriums when there are only two files in the system. The prices of anarchy, both from each node's perspective and the whole system's perspective, are also characterized. Moreover, we propose a distributed mechanism where each node adjusts its strategies only based on local information and show that the mechanism converges to a Nash Equilibrium. We also introduce an approach for calculating Nash Equilibriums for systems that incorporate network coding when there are more than two files.
\end{abstract}

\section{Introduction}
\label{section:introduction}

It is known that using wireless peer-to-peer (P2P) networks to distribute information locally can improve the system performance in many different aspects. For example, in cellular networks, nearby mobile devices can exchange data with each other instead of contacting the faraway base station for the data. Since exchanging data with nearby devices requires much less power and results in less interference to other devices, such an approach may reduce power consumption and increase spatial reuse. 

Many existing studies (e.g., \cite{HH04,MN11, JZ08}) have demonstrated the benefits of wireless P2P networks. However, these studies have assumed that all nodes are cooperative. In practice, nodes may be selfish. Therefore, a major challenge for P2P networks is to provide incentives to nodes in the network so that they are willing to contribute to the network by transmitting data that they possess. While there are a lot of studies, such as \cite{CA11,SL09, VM10}, on this topic for P2P networks over Internet, these works cannot be applied to wireless P2P networks. Due to the broadcast nature of wireless transmissions, when a node transmits a packet, all nodes within the proximity are able to receive the packet. Therefore, in wireless P2P networks, data exchange involves all nodes within the system, rather than only two nodes as in P2P networks over Internet.

In this paper, we study wireless P2P networks composed of selfish nodes. We first provide a model that considers the broadcast nature of wireless transmissions and the incentives of selfish nodes. Each node in the system aims to increase its download rate and decrease its upload rate, so as to reduce its own power consumption. We then propose a protocol for content distribution in these wireless P2P networks. Our protocol does not require the exchange of money, reputation, etc., and hence can be implemented without the need of additional infrastructure. This non-monetary feature further distinguishes our work from other studies that rely on additional infrastructure to set prices or payoffs \cite{JP10, VM10, ZH09}, or to punish uncooperative nodes \cite{WL11}. Moreover, our protocol can be easily modified to employ network coding. 

The performance of our protocol is studied. When there are only two files in the system, we show that there are closed-form expressions for each node's strategies under a series of Nash Equilibriums. We also derive the prices of anarchy under these Nash Equilibriums, both from a node's selfish perspective and the whole system's perspective.

To compute its strategy under a Nash Equilibrium, a node needs information of all other nodes, which is not always available to the node. To address this challenge, we propose a distributed mechanism where each node updates its strategy only based on its private information and the history of the system. We show that this distributed mechanism converges to a Nash Equilibrium. Moreover, this mechanism is also consistent with each node's incentive, as the expected cost of each node reduces with each update.

We then consider systems that have more than two files and employ network coding. We propose a systematic approach to compute the Nash Equilibriums. The performance of such systems are further investigated through numerical studies.

The rest of the paper is organized as follows. Section \ref{section:protocol} proposes our system model and protocol for content distribution. Section \ref{section:bilateral} studies the Nash Equilibriums for systems with only two files. Section \ref{section:price of anarchy} studies the prices of anarchy under these Nash Equilibriums. Section \ref{section:implementation} discusses implementation issues and provides a distributed mechanism for nodes to update their strategies. Section \ref{section:network coding} studies the Nash Equilibriums when there are more than two files in systems that employ network coding. Section \ref{section:numerical} provides some numerical results. Finally, Section \ref{section:conclusion} concludes the paper.

\section{System Model and Protocol Overview}
\label{section:protocol}

Consider a system with a number of wireless nodes that share a set of files or video/audio streams. We assume that each file, or stream, is composed of infinitely many packets. This assumption is natural when nodes share a set of streams. In scenarios where nodes share a set of files, this assumption is a good approximation when the size of each file is at least several MBytes, or, consists of at least thousands of packets. We denote the set of nodes by $N$, and the set of files, or streams, by $X=\{A,B,C,\dots\}$. We assume that each node needs to download one file from other peers, and holds all packets of all the other files. One way to enforce this assumption is by requiring each node to download all but one files through, say, its own 3G connection before admitting the node into the system. We use $X_n$ to denote the file that client needs. For example, if $X_n=B$, then node $n$ needs to download $B$ from its peers, and have all data of files $A,C,D,\dots.$ In this paper, we use upper-case letters to represent files and lower-case letters to represent nodes. When the file that a node needs is not important in the context, we use $n,m,$ etc., to denote a node. On the other hand, when we want to stress that the indicated node needs a certain file, say, file $A$, we use $a_i$ to denote the node.

%We use $X_n=[Y_n|Z_n]$ to denote the set of files $Y_n$ that node $n$ needs and the set of files $Z_n$ that node $n$ possesses. For example, if $X_n=[A,B|C,D]$, then node $n$ holds every data of files $C$ and $D$, and node $n$ needs other peers to transmit every data of files $A$ and $B$ to it. In this paper, we use upper-case letters to represent files and lower-case letters to represent nodes. When the file that a node needs is not important in the context, we use $n,m,$ etc., to denote a node. On the other hand, when we want to stress that the indicated node needs a certain file, say, file $A$, we use $a_i$ to denote the node.

Most protocols for peer-to-peer content distribution involves the data exchange between two peers. However, those protocols may not be directly applicable to wireless broadcast networks. In wireless networks, all transmissions are carried by the shared wireless medium. We assume that, when a node transmits a packet from a file, all nodes receive this packet. The broadcasting nature of wireless transmissions may result in a ``\emph{free rider}'' problem as depicted in the following example.

\begin{example}
Consider a system with four nodes and two files where $X_1=X_2=A$ and $X_3=X_4=B$. Suppose that node 1 and node 3 exchange data, that is, node 1 transmits packets of $B$, and node 3 transmits packets of $A$ in return. As all nodes can receive all transmissions, node 2 and node 4 can obtain packets of $A$ and $B$ without making any transmissions. Therefore, we say that node 2 and node 4 are \emph{free riders}.
\end{example}

In addition to being unfair, the presence of free riders may prevent nodes from transmitting data and contributing to the networks when all nodes are selfish. In the above example, each node may refrain from transmitting data, in the hope that other nodes participate in exchanging data, making itself a free rider.

In this paper, we propose a non-monetary protocol for P2P content distribution and study its performance when all nodes are selfish. Before introducing the protocol, we first formally describe the goal of each node.

We define the time needed for transmitting a data packet to be one unit time. We assume that the goal of a node is to increase its download rate and to reduce the number of transmissions it makes. To be more specific, we assume that whenever node $n$ transmits a data packet, it needs to pay a \emph{transmission cost} of $g_n$. The transmission cost can be chosen to, for example, reflect the amount of power needed for making a transmission. On the other hand, node $n$ pays a \emph{waiting cost} at rate $w_n$ per unit time through the course of the protocol. Suppose that, a node $n$ receives two needed packets at time $t_1$ and $t_2$, respectively. During time $[t_1,t_2]$, node $n$ transmits a total number of $\alpha_n$ packets. Thus, to download the packet at time $t_2$, node $n$ waits a total amount of $t_2-t_1$ time, and makes $\alpha_n$ transmissions. The \emph{total cost} for the packet that node $n$ downloads at $t_2$ can then be expressed as $\alpha_ng_n+(t_2-t_1)w_n$. The \emph{average total cost} of node $n$ is then defined as the long-term average total cost per downloaded packet. The following example explains the computation of average total cost.

\begin{example}
Suppose a node $n$ downloads a total number of $d_n(t)$ packets that it needs, and transmits $u_n(t)$ packets, during time $[0,t]$. The average total cost of node $n$ is then $\limsup_{t\rightarrow\infty}\frac{u_n(t)g_n+tw_n}{d_n(t)}$.
\end{example}

We assume that the goal of node $n$ is to minimize its average total cost. Node $n$ wishes to reduce $\limsup_{t\rightarrow\infty}\frac{u_n(t)}{d_n(t)}$ in order to achieve small transmission cost. On the other hand, the download rate of node $n$ can be expressed as $\liminf_{t\rightarrow\infty}\frac{d_n(t)}{t}$, which is the inverse of $\limsup_{t\rightarrow\infty}\frac{t}{d_n(t)}$. Therefore, to have high download rate, node $n$ also wishes to have small $\limsup_{t\rightarrow\infty}\frac{t}{d_n(t)}$.

We now describe our protocol for P2P content distribution. The process is composed of \emph{rounds}, where  two nodes transmit two packets that each other needs in each round. At the beginning of a round, each node $n$ secretly and randomly picks a backoff timer, $t_n$. Node $n$ then waits and listens to the channel for $t_n$ time. If no transmissions take place in $t_n$ time, node $n$ transmits a control packet that contains its value of $X_n$. This control packet is interpreted as a promise from node $n$ saying that if any node transmits a packet of $X_n$, node $n$ will respond with a packet that $n$ has. As the length of this control packet is much smaller than a data packet, we assume that time needed for transmitting the control packet is negligible and node $n$ does not pay transmission cost for the control packet.  After node $n$ transmits the control packet, every node $m$ that has the file $X_n$ secretly and randomly picks a backoff timer $\hat{t}_m$. Node $m$ waits and listens to the channel for $\hat{t}_m$ time. If no other nodes transmit in $\hat{t}_m$ time, node $m$ transmits a data packet of $X_n$, and piggybacks its value of $X_m$. Upon receiving the data packet from node $m$, node $n$ responds with a packet of $X_m$, as promised in its control packet. The round ends after node $n$ responds the packet and a new round begins.

\begin{figure}[t]
\includegraphics[width = 3.3in]{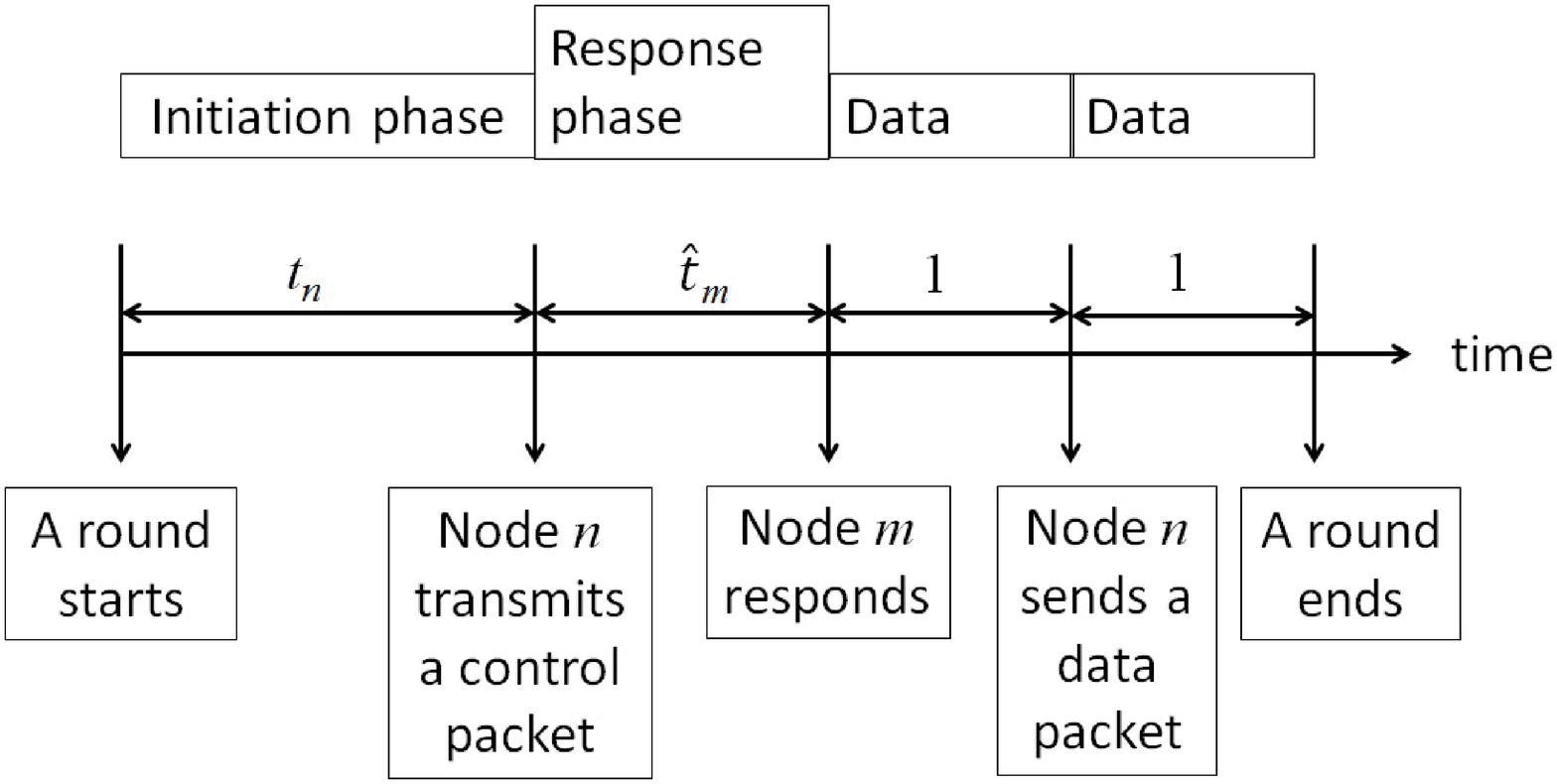}
\caption{An example of a round.} \label{fig:model:example}
\end{figure}

As explained in the previous paragraph, there are two phases of backoffs in a round, one for a node to transmit a control packet, and the other for a node to respond to the control packet by transmitting a data packet. We call the two phases of backoffs the \emph{initiation phase} and the \emph{response phase}, respectively. Figure \ref{fig:model:example} shows an example of a round.

Intuitively, when a node $n$ chooses a large value of $t_n$ and $\hat{t}_n$, it is likely that node $n$ does not transmit, which increases its chance of being a free rider and reduces its transmission cost. However, a large value of $t_n$ and $\hat{t}_n$ also means that node $n$ may need to wait a long time before receiving a packet it needs, which results in large waiting cost. By taking waiting costs into account, our protocol encourages nodes to choose reasonable small value of $t_n$ and $\hat{t}_n$, and hence enables data exchange among nodes. We also note that this protocol is non-monetary and can be easily implemented for modern wireless networks without the need of additional infrastructure.

Finally, we show that our protocol can be modified to incorporate network coding. The modified protocol is very similar to the one shown in Fig. \ref{fig:model:example}. The only difference is that, in the last part of a round, when node $n$ sends a data packet, it sends a coded packet that contains one packet of each file that it has. For example, suppose that there are three files $A,B,C$ in the system, and $X_n=A$, then client $n$ sends a data packet containing (one packet of $B$)+(one packet of $C$) after it receives a packet of $A$ from $m$. Since we assume that every node possesses all but one files, the coded packet sent by $n$ can be decoded by every client that needs either file $B$ or file $C$. By using network coding, we guarantee that every client in the system obtains one packet in each round. Also, since clients that possess neither file $B$ nor file $C$ cannot decode the packet sent by $n$, this approach also prevents clients that need more than one files from joining the system. 

\subsection{Performance Evaluation}

Under our protocol, each selfish node's strategy consists of two parts: choosing a distribution to generate its backoff timer in the initiation phase, $t_n$, and choosing a distribution to generate its backoff timer in the response phase, $\hat{t}_n$. We say that the strategies of all nodes in the system form a \emph{Nash Equilibrium} if, for each node $n$, its strategy minimizes its average total cost, given the strategies of all other nodes.

In the following sections, we analyze the performance of our protocol under Nash Equilibriums. We consider the performance from both nodes' perspectives and the system's perspective. A node's performance is simply based on its average total cost. On the other hand, we consider the system's performance by its per-node average throughput, defined as the average download rate over all clients. We also define the \emph{price of anarchy on node cost} and \emph{price of anarchy on system throughput} as follows:

\begin{definition}
The \emph{price of anarchy on node cost} of a node $n$ under a Nash Equilibrium is the ratio of the average total cost of $n$ under the Nash Equilibrium and the minimum possible average total cost of $n$ when all nodes are cooperative.
\end{definition}

\begin{definition}
The \emph{price of anarchy on system throughput} under a Nash Equilibrium is defined as $$\frac{(\mbox{maximum possible per-node throughput})}{(\mbox{per-node throughput under the Nash Equilibrium})}.$$
\end{definition}

%\section{Pure Strategy}

%In this section, we will show that for a one-shot game, that is, the file consists of only one packet, the only pure strategy Nash Equilibrium is $t_n=0$ for one $n$, and $t_m=\infty$ for all others. This equilibrium is extremely unfair and is not stable. (We will need a proper definition for stable)

\section{Nash Equilibriums for Bilateral File Exchanges}	\label{section:bilateral}

In this section, we analyze the performance of our protocol when there are only two files in the system. In such a system, we can divide nodes into two groups, nodes in one group, indexed by $a_1,a_2,\dots, a_I$, have the file $B$ and need the file $A$, that is, $X_{a_i}=A$, while nodes in the other group, indexed by $b_1, b_2,\dots, b_J$ have $X_{b_j}=B$. In this setting, network coding is not employed.

We will show that there is a series of Nash Equilibriums where each node $n$ chooses each of its backoff timers in the initiation phase and response phase as an exponential random variable. We focus on exponential random variables due to its memoryless property, which makes it easily implementable. 

%By settingSecond, as the strategy of each node is the same in each round, a node can minimize its average total cost by greedily minimizing the expected total cost for each packet it downloads. Hence, throughout the rest of this section, we will study the expected total cost for each node in a round.

Our analysis is based on the following theorem:

\begin{theorem}	\label{theorem:condition for NE} \cite[Lemma~3.1]{textbook}
The strategies of nodes in a game form a Nash Equilibrium if, for each node $n$, given the strategies of other nodes, the expected cost of node $n$ is the same regardless of its chosen values of backoff timers.
\end{theorem}

We first consider the nodes' strategies on choosing the backoff timers in the response phase, $\hat{t}_n$. Without loss of generality, we assume that node $b_1$ has sent the control packet in the initiation phase. Each node $a_i$ secretly and randomly chooses a backoff timer $\hat{t}_{a_i}$. Assume that the timer chosen by node $a_{i^*}$ is the smallest among all backoff timers, that is, $\hat{t}_{a_{i^*}}=\min\{\hat{t}_{a_1}, \hat{t}_{a_2},\dots\}$. Node $a_{i^*}$ will transmit after time $\hat{t}_{a_{i^*}}$, and the \emph{additional waiting time}, which is the length of the response phase, for each node $a_i$ to download a packet is $\hat{t}_{a_{i^*}}$. In the expression of the additional waiting time, we exclude the time needed waiting for $b_1$ to transmit its control packet and the time needed for transmitting data packets, as these times are not influenced by the values of $\hat{t}_{a_i}$. In the following, we also exclude the waiting cost incurred by the time waiting for $b_1$ to transmit its control packet and the time needed for transmitting data packets when discussing \emph{additional waiting cost} and \emph{additional total cost}.

We show that there is a Nash Equilibrium where each node $a_i$ chooses its backoff timer as an exponential random variable. In particular, we assume that $\hat{t}_{a_i}\sim EXP(\lambda_{a_i})$, where the value of $\lambda_{a_i}$ will be determined in the sequel.

We now apply Theorem \ref{theorem:condition for NE} to determine the values of $\lambda_{a_i}$. Suppose that node $a_{i^*}$ chooses $\hat{t}_{a_{i^*}}=t$. If $t$ is smaller than all $\hat{t}_{a_i}$, $i\neq i^*$, that is, $t<\min_{i\neq i^*}\hat{t}_{a_i}$, node $a_{i^*}$ needs to transmit a packet after waiting for $t$ units of time, and thus the additional total cost for node $a_{i^*}$ on this packet is $g_{a_{i^*}}+w_{a_{i^*}}t$. On the other hand, if $t>\min_{i\neq i^*}\hat{t}_{a_i}$, a node other than $a_{i^*}$ transmits the packet, and node $a_{i^*}$ becomes the free rider. In this case, the additional total cost of node $a_{i^*}$ on this packet is $w_{a_{i^*}}(\min_{i\neq i^*}\{\hat{t}_{a_i}\})$. Note that we have $\min_{i\neq i^*}\{\hat{t}_{a_i}\}\sim EXP(\sum_{i\neq i^*}\lambda_{a_i})$. Let $\lambda_{-i^*}:=\sum_{i\neq i^*}\lambda_{a_i}$. The expected additional total cost of node $i^*$ can be written as:

\begin{align}
&\int_{s=0}^t w_{a_{i^*}}s\lambda_{-i^*}e^{-\lambda_{-i^*}s}ds\nonumber\\
&+\int_{s=t}^\infty(g_{a_{i^*}}+w_{a_{i^*}}t)\lambda_{-i^*}e^{-\lambda_{-i^*}s}ds\nonumber\\
%=&\frac{w_{a_{i^*}}}{\lambda_{-i^*}}+e^{-\lambda_{-i^*}t}(-w_{a_{i^*}}t-\frac{w_{a_{i^*}}}{\lambda_{-i^*}})\nonumber\\
%&+(w_{a_{i^*}}t+g_{a_{i^*}})e^{-\lambda_{-i^*}t}\nonumber\\
=&\frac{w_{a_{i^*}}}{\lambda_{-i^*}}+e^{-\lambda_{-i^*}t}(g_{a_{i^*}}-\frac{w_{a_{i^*}}}{\lambda_{-i^*}}). \label{equation:bilateral:costA}
\end{align}

Thus, if $g_{a_i}=\frac{w_{a_i}}{\lambda_{-i}}$, for all $i$, the strategies form a Nash Equilibrium. This can be done by choosing
\begin{equation}	\label{equation:bilateral:lambdaA}
\lambda_{a_i}=\frac{1}{I-1}\sum_{k}\frac{w_{a_k}}{g_{a_k}}-\frac{w_{a_i}}{g_{a_i}},
\end{equation}
where $I$ is the number of nodes in group $\{a_1,a_2,\dots\}$. We can also conclude that the expected value of $\min_i\{\hat{t}_{a_i}\}$, which is the expected duration of the response phase, given that a node in the group $\{b_1,b_2,\dots\}$ transmits the control packet, is
\begin{equation}	\label{equation:bilateral:hatTA}
\hat{T}_A:=\frac{1}{\sum_{i}\lambda_{a_i}}=\frac{I-1}{\sum_i w_{a_i}/g_{a_i}},
\end{equation}
and that the expected additional total cost of node $a_i$ is $g_{a_i}$.

Similarly, if a node $a_i$ transmits a control packet, each node $b_j$ selects $\hat{t}_{b_j}\sim EXP(\lambda_{b_j})$, where
\begin{equation}	\label{equation:bilateral:lambdaB}
\lambda_{b_j}=\frac{1}{J-1}\sum_{k}\frac{w_{b_k}}{g_{b_k}}-\frac{w_{b_j}}{g_{b_j}}.
\end{equation}
The expected duration of the response phase is
\begin{equation}\label{equation:bilateral:hatTB}
\hat{T}_B:=\frac{1}{\sum_{j}\lambda_{b_j}}=\frac{J-1}{\sum_j w_{b_j}/g_{b_j}},
\end{equation}
and the expected additional total cost of node $b_j$ is $g_{b_j}$.

Next, we consider the choice of backoff timer in the initiation phase, that is, the choice of $t_{n}$ for a node $n$. We will show that there is a Nash Equilibrium where each node $a_i$ selects $t_{a_i}\sim EXP(\gamma_{a_i})$ and each node $b_j$ selects $t_{b_j}\sim EXP(\gamma_{b_j})$.

Assume that a node, say, node $a_{i^*}$, selects $t_{a_{i^*}}=t$. If $t$ is the smallest timer among all timers, that is, $t<\min_{i\neq i^*}t_{a_i}$ and $t<\min_jt_{b_j}$, node $a_{i^*}$ transmits the control packet after time $t$. After which time, it needs to wait one of the nodes in $\{b_1,b_2,\dots\}$ to respond with a data packet, and then $a_{i^*}$ needs to transmit a data packet. By the analysis above, we know that the expected time that $a_{i^*}$ waits for one of the nodes in $\{b_1,b_2,\dots\}$ to respond is $\hat{T}_B$. Thus, the expected total cost for node $a_{i^*}$ is $g_{a_{i^*}}+w_{a_{i^*}}(t+\hat{T}_B)+2w_{a_{i^*}}$, where the last term, $2w_{a_{i^*}}$, accounts for the waiting cost caused by transmission delays, as it takes two units time to transmit two data packets.

Next, consider the case that $t>\min\{\min_{i\neq i^*}t_{a_i}, \min_jt_{b_j}\}$. We have that $\min_{i\neq i^*}t_{a_i}\sim EXP(\sum_{i\neq i^*}\gamma_{a_i})$ and $\min_jt_{b_j}\sim EXP(\sum_j \gamma_{b_j})$. By the memoryless property of exponential functions, we have that
\begin{align*}
&P_{A\backslash\{i^*\}<B}\\
:=&Prob\{\min_{i\neq i^*}t_{a_i} < \min_jt_{b_j}|t>\min\{\min_{i\neq i^*}t_{a_i}, \min_jt_{b_j}\}\}\\
=&\frac{\sum_{i\neq i^*}\gamma_{a_i}}{\sum_{i\neq i^*}\gamma_{a_i}+\sum_{j}\gamma_{b_j}}.
\end{align*}
That is, with probability $P_{A\backslash\{i^*\}<B}$, one of the nodes in $\{a_1,a_2,\dots\}$ other than $a_{i^*}$ transmits the control packet, and ,with probability $1-P_{A\backslash\{i^*\}<B}$, one of the nodes in $\{b_1,b_2,\dots\}$ transmits the control packet. If it is the former case, node $a_{i^*}$ does not need to transmit any packets, and its expected cost is $w_{a_{i^*}}(\min\{\min_{i\neq i^*}t_{a_i}, \min_jt_{b_j}\}+\hat{T}_B+2)$. If it is the later case, the expected cost is $w_{a_{i^*}}(\min\{\min_{i\neq i^*}t_{a_i}, \min_jt_{b_j}\}+2)+g_{a_{i^*}}$, since we have shown that the expected additional total cost of $a_{i^*}$ is $g_{a_{i^*}}$. Hence, given that $t>\min\{\min_{i\neq i^*}t_{a_i}, \min_jt_{b_j}\}$, the expected cost is $w_{a_{i^*}}(\min\{\min_{i\neq i^*}t_{a_i}, \min_jt_{b_j}\}+2)+P_{A\backslash\{i^*\}<B}w_{a_{i^*}}\hat{T}_B+(1-P_{A\backslash\{i^*\}<B})g_{a_{i^*}}$.

As we have $\min\{\min_{i\neq i^*}t_{a_i}, \min_jt_{b_j}\}\sim EXP(\sum_{i\neq i^*}\gamma_{a_i}+\sum_j\gamma_{b_j})$, by letting $\gamma_{-i^*}:=\sum_{i\neq i^*}\gamma_{a_i}+\sum_j\gamma_{b_j}$, the expected cost of $a_{i^*}$ can be computed as:
\begin{align}
&\int_{s=0}^t [w_{a_{i^*}}(s+2)+P_{A\backslash\{i^*\}<B}w_{a_{i^*}}\hat{T}_B]\gamma_{-i^*}e^{-\gamma_{-i^*}s}ds\nonumber\\
&+\int_{s=0}^t [(1-P_{A\backslash\{i^*\}<B})g_{a_{i^*}}]\gamma_{-i^*}e^{-\gamma_{-i^*}s}ds\nonumber\\
&+\int_{s=t}^\infty (g_{a_{i^*}}+w_{a_{i^*}}(t+\hat{T}_B)+2w_{a_{i^*}})\gamma_{-i^*}e^{-\gamma_{-i^*}s}ds\nonumber\\
%=&\frac{w_{a_{i^*}}}{\gamma_{-i^*}}-e^{-\gamma_{-i^*}t}(\frac{w_{a_{i^*}}}{\gamma_{-i^*}}+w_{a_{i^*}}t)\nonumber\\
%&+(1-e^{-\gamma_{-i^*}t})(P_{A\backslash\{i^*\}<B}w_{a_{i^*}}\hat{T}_B+(1-P_{A\backslash\{i^*\}<B})g_{a_{i^*}})\nonumber\\
%&+e^{-\gamma_{-i^*}t}(g_{a_{i^*}}+w_{a_{i^*}}(t+\hat{T}_B))+2w_{a_{i^*}}\nonumber\\
=&\frac{w_{a_{i^*}}}{\gamma_{-i^*}}+\frac{\sum_{i\neq i^*}\gamma_{a_i}}{\gamma_{-i^*}}
w_{a_{i^*}}\hat{T}_B+\frac{\sum_{j}\gamma_{b_j}}{\gamma_{-i^*}}g_{a_{i^*}}+2w_{a_{i^*}}\nonumber\\
+&e^{-\gamma_{-i^*}t}(\frac{\sum_{j}\gamma_{b_j}}{\gamma_{-i^*}}w_{a_{i^*}}\hat{T}_B+\frac{\sum_{i\neq i^*}\gamma_{a_i}}{\gamma_{-i^*}}g_{a_{i^*}}-\frac{w_{a_{i^*}}}{\gamma_{-i^*}}).\label{equation:bilateral:total cost}
\end{align}

We wish to find $\{\gamma_{a_1}, \gamma_{a_2},\dots, \gamma_{b_1},\gamma_{b_2},\dots\}$ so that the expected cost of $a_{i^*}$ is the same for all $t$. Hence, we require that
\begin{align}
&(1-P_{A\backslash\{i^*\}<B})w_{a_{i^*}}\hat{T}_B+P_{A\backslash\{i^*\}<B}g_{a_{i^*}}=\frac{w_{a_{i^*}}}{\gamma_{-i^*}}\\
&\Leftrightarrow (\sum_{j}\gamma_{b_j})\hat{T}_{B}w_{a_{i^*}}+(\sum_{i\neq i^*}\gamma_{a_i})g_{a_{i^*}}=w_{a_{i^*}}\\
&\Leftrightarrow \frac{w_{a_{i^*}}}{g_{a_{i^*}}}=\frac{\sum_{i\neq i^*}\gamma_{a_i}}{1-(\sum_{j}\gamma_{b_j})\hat{T}_B},	\label{equation:bilateral:eqA}
\end{align}
for all $a_{i^*}$. Similarly, by studying the expected cost of a node $b_{j^*}$, we also require that
\begin{equation}
\frac{w_{b_{j^*}}}{g_{b_{j^*}}}=\frac{\sum_{j\neq j^*}\gamma_{b_j}}{1-(\sum_{i}\gamma_{a_i})\hat{T}_A}, \label{equation:bilateral:eqB}
\end{equation}
for all $b_{j^*}$.

Summing the (\ref{equation:bilateral:eqA}) over all $a_{i^*}$ and we have
\begin{align}
&\sum_{i}\frac{w_{a_{i}}}{g_{a_{i}}}=\frac{(I-1)\sum_{i}\gamma_{a_i}}{1-(\sum_{j}\gamma_{b_j})\hat{T}_B}	 \label{equation:bilateral:eqAA}\\
\Leftrightarrow &(\sum_{i}\gamma_{a_j})\hat{T}_A+(\sum_{j}\gamma_{b_j})\hat{T}_B=1.	 \label{equation:bilateral:sum}
\end{align}

Assume that $(\sum_{i}\gamma_{a_j})\hat{T}_A=\alpha$ and $(\sum_{j}\gamma_{b_j})\hat{T}_B=1-\alpha$, for some $\alpha\in(0,1)$. Using (\ref{equation:bilateral:eqA}) and (\ref{equation:bilateral:eqAA}), we obtain
\begin{align}
\gamma_{a_{i^*}}=\sum_{i}\gamma_{a_i}-\sum_{i\neq i^*}\gamma_{a_i}=\alpha(\frac{\sum_iw_{a_i}/g_{a_i}}{I-1}-\frac{w_{a_{i^*}}}{g_{a_{i^*}}}).	 \label{equation:bilateral:solA}
\end{align}
Similarly, we also obtain
\begin{align}
\gamma_{b_{j^*}}=(1-\alpha)(\frac{\sum_jw_{b_j}/g_{b_j}}{J-1}-\frac{w_{b_{j^*}}}{g_{b_{j^*}}}). \label{equation:bilateral:solB}
\end{align}

It is easy to check that, for every $\alpha\in(0,1)$, setting $\gamma_{a_i}$ and $\gamma_{b_j}$ according to (\ref{equation:bilateral:solA}) and (\ref{equation:bilateral:solB}) satisfies (\ref{equation:bilateral:eqA}) and (\ref{equation:bilateral:eqB}) for all nodes, and the expected cost of each node is the same regardless the actual backoff timer it chooses. Hence, (\ref{equation:bilateral:solA}) and (\ref{equation:bilateral:solB}) form a Nash Equilibrium. Further, as $\alpha$ can be any number in $(0,1)$, this game has infinitely many Nash Equilibriums.

We summarize our results for systems with only two files in the following theorem.
\begin{theorem}	\label{theorem:bilateral:NE}
For any $\alpha\in(0,1)$, if each node $a_{i^*}$ chooses $t_{a_{i^*}}\sim EXP(\gamma_{a_{i^*}})$ and $\hat{t}_{a_{i^*}}\sim EXP(\lambda_{a_{i^*}})$, and each node $b_{j^*}$ chooses $t_{b_{j^*}}\sim EXP(\beta_{b_{j^*}})$ and $\hat{t}_{b_{j^*}}\sim EXP(\lambda_{b_{j^*}})$, where $\gamma_{a_{i^*}}, \lambda_{a_{i^*}}, \gamma_{b_{j^*}}$, and $\lambda_{b_{j^*}}$ are chosen by (\ref{equation:bilateral:solA}), (\ref{equation:bilateral:lambdaA}),  (\ref{equation:bilateral:solB}), and (\ref{equation:bilateral:lambdaB}), then these strategies form a Nash Equilibrium.
\end{theorem}

%In the above theorem, each node $n$ needs to have global knowledge of the whole network to compute the values of $\lambda_n$ and $\gamma_n$, and hence it may be infeasible to implement these policies in a distributed fashion. We now propose
%(\textbf{We probably should include some sentences on why we consider exponential distributions here.})

\section{Price of Anarchy for Bilateral File Exchanges}
\label{section:price of anarchy}

We have found Nash Equilibriums for our protocol when there are only two files in the system. We now discuss the performances of these Nash Equilibriums.

Suppose all nodes choose their backoff timers according to Theorem \ref{theorem:bilateral:NE}, for some $\alpha\in(0,1)$. The duration of the initiation phase is $\min\{t_{a_1}, t_{a_2},\dots, t_{b_1}, t_{b_2},\dots\}$, which is an exponential random variable with mean
\begin{equation}
T_{A,B}:=\frac{1}{\sum_{i}\gamma_{a_i}+\sum_j\gamma_{b_j}}.
\end{equation}
Also, the probability that one of the nodes in group $\{a_1,a_2,\dots\}$ transmits the control packet is
\begin{equation}
Prob\{\min_it_{a_i}<\min_jt_{b_j}\}=\frac{\sum_i\gamma_{a_i}}{\sum_{i}\gamma_{a_i}+\sum_j\gamma_{b_j}}.
\end{equation}

Using (\ref{equation:bilateral:hatTA}) and (\ref{equation:bilateral:hatTB}), we can express the expected amount of time for the two groups of nodes to exchange two packets as
\begin{align}
&T_{A,B}+\frac{\sum_i\gamma_{a_i}}{\sum_{i}\gamma_{a_i}+\sum_j\gamma_{b_j}}\hat{T}_B+\frac{\sum_j\gamma_{b_j}}{\sum_{i}\gamma_{a_i}+\sum_j\gamma_{b_j}}\hat{T}_A+2\nonumber\\
%=&\frac{1+\sum_i\gamma_{a_i}\hat{T}_B+\sum_j\gamma_{b_j}\hat{T}_A}{\sum_{i}\gamma_{a_i}+\sum_j\gamma_{b_j}}+2\nonumber\\
%=&\frac{(\sum_i\gamma_{a_i}+\sum_j\gamma_{b_j})(\hat{T}_B+\hat{T}_A)}{\sum_{i}\gamma_{a_i}+\sum_j\gamma_{b_j}}+2 \hspace{30pt}\mbox{(by (\ref{equation:bilateral:sum}))}\nonumber\\
&=\hat{T}_A+\hat{T}_B+2,
\end{align}
where the last term in the equation accounts for the time needed for transmitting two packets. Since every node downloads a packet in each round, the per-node average throughput of our protocol is then $1/(\hat{T}_A+\hat{T}_B+2)$ packets per unit time under the described Nash Equilibriums. On the other hand, the download rate of a client is at most 0.5 packet per unit time, and the maximum possible per-node throughput is 0.5. Therefore, the price of anarchy on system throughput is $(\hat{T}_A+\hat{T}_B+2)/2$. To better understand the price of anarchy on system throughput, we consider the special case where all nodes have the same parameters for waiting cost and for transmission cost, that is, $w_n\equiv w$ and $g_n\equiv g$ for all $n$. In this case, we have $\hat{T}_A=\frac{(I-1)(g/w)}{I}$ and $\hat{T}_B=\frac{(J-1)(g/w)}{J}$, and the price of anarchy on system throughput is $(\frac{(I-1)(g/w)}{I}+\frac{(J-1)(g/w)}{J}+2)/2\leq 1+\frac{g}{w}$.

We now compute the average total costs of nodes. The probability that a node $a_{i^*}$ transmits a packet in a round can be expressed as
\begin{align*}
&Prob\{\min_it_{a_i}<\min_jt_{b_j}\}Prob\{t_{a_{i^*}}<t_{a_i},\forall i\neq i^*\}\\
&+Prob\{\min_it_{a_i}>\min_jt_{b_j}\}Prob\{\hat{t}_{a_{i^*}}<\hat{t}_{a_i},\forall i\neq i^*\}\\
%=&Prob\{\min_it_{a_i}<\min_jt_{b_j}\}\frac{\lambda_{a_{i^*}}}{\sum_i\lambda_{a_i}}\\
%&+Prob\{\min_it_{a_i}>\min_jt_{b_j}\}\frac{\gamma_{a_{i^*}}}{\sum_i\gamma_{a_i}}\\
=&(\frac{\sum_iw_{a_i}/g_{a_i}}{I-1}-\frac{w_{a_{i^*}}}{g_{a_{i^*}}})/(\frac{\sum_iw_{a_i}/g_{a_i}}{I-1}).
\end{align*}
The average total cost of node $a_{i^*}$ is then
\begin{align}
&w_{a_{i^*}}(\hat{T}_A+\hat{T}_B+2)\nonumber\\
&+g_{a_{i^*}}(\frac{\sum_iw_{a_i}/g_{a_i}}{I-1}-\frac{w_{a_{i^*}}}{g_{a_{i^*}}})/(\frac{\sum_iw_{a_i}/g_{a_i}}{I-1})\nonumber\\
=&g_{a_{i^*}}+\frac{(J-1)w_{a_{i^*}}}{\sum_jw_{b_j}/g_{b_j}}+2w_{a_{i^*}}.
\end{align}

On the other hand, under our protocol, the download rate of node $a_{i^*}$ is at most one packet per 2 unit times. Hence, the average total cost of node $a_{i^*}$ is at least $2w_{a_{i^*}}$. We then have that the price of anarchy on node cost of $a_{i^*}$ is at most $(\frac{g_{a_{i^*}}}{w_{a_{i^*}}}+\frac{(J-1)}{\sum_jw_{b_j}/g_{b_j}}+2)/2$. In the special case where $w_n\equiv w$ and $g_n\equiv g$, for all nodes,, the price of anarchy on node cost of $a_{i^*}$ is at most $(\frac{g}{w}+\frac{(J-1)(g/w)}{J}+2)/2\leq 1+\frac{g}{w}.$

We note that both the price of anarchy on system throughput and the price of anarchy on node cost increase with $\frac{g}{w}$. Intuitively, when $g$ is small compared to $w$, nodes focus more on improving their download rate than on reducing transmission cost. Hence, nodes tend to choose small backoff timers and result in small price of anarchy. On the other hand, when $g$ is much larger than $w$, transmission costs become an important factor of nodes' costs. Hence, each node tends to choose large backoff timers to increase its chance of becoming a free rider, which results in large price of anarchy.

As a final remark, we note that both the system throughput and total average costs of nodes remain the same for all $\alpha\in(0,1)$. Therefore, the performance of the system is the same for all Nash Equilibriums described by Theorem \ref{theorem:bilateral:NE}.

%\textbf{We need to formally define price of anarchy somewhere}

\section{Implementation Issues and Convergence}	\label{section:implementation}

Section \ref{section:bilateral} has described a Nash Equilibrium for a system with two files. However, for a node $n$ to derive its strategies, that is, to compute the values of $\gamma_n$ and $\lambda_n$, node $n$ needs to know information of the whole network, including the private values of $w_m$ and $g_m$ for all other nodes $m$. In this section, we propose a distributed mechanism for each node to update its values of $\gamma_n$ and $\lambda_n$ only based on its values of $w_n$ and $g_n$ and the history of the system. We show that this mechanism is compatible to the node's incentive, in the sense that the updated $\gamma_n$ and $\lambda_n$ achieve smaller average total cost for the node. Moreover, we also show that the system converges to a Nash Equilibrium when all nodes apply this mechanism.

We order the two flies by lexicographical order. If file $A$ has higher order than $B$, we impose that $\gamma_{a_i}=0$, for all $i$, and $\lambda_{b_j}=0$, for all $j$. This corresponds to the case where $\alpha=0$ in Section \ref{section:bilateral}. Therefore, in every round, a node in $\{b_1,b_2,\dots\}$ transmits a control packet in the initiation phase and a node in $\{a_1,a_2,\dots\}$ transmits a data packet in the response phase. On the other hand, if file $B$ has higher order than $A$, we impose that $\lambda_{a_i}=0$, for all $i$, and $\gamma_{b_j}=0$, for all $j$, which corresponds to the case where $\alpha=1$. Without loss of generality, we assume that file $A$ has higher order than $B$.

Using (\ref{equation:bilateral:lambdaA}), (\ref{equation:bilateral:hatTA}), (\ref{equation:bilateral:solB}) , we have that $\hat{T}_A=\frac{1}{\sum_i\lambda_{a_i}}=\frac{I-1}{\sum_i\frac{w_{a_i}}{g_{a_i}}}$, $\lambda_{a_{i^*}}=\frac{1}{\hat{T}_A}-\frac{w_{a_{i^*}}}{g_{a_{i^*}}}$, $\hat{T}_B= \frac{J-1}{\sum_j\frac{w_{b_j}}{g_{b_j}}}$, and $\gamma_{b_{j^*}}=\frac{1}{\hat{T}_B}-\frac{w_{b_{j^*}}}{g_{b_{j^*}}}$ at the Nash Equilibrium, where $I$ and $J$ are the number of nodes in groups $\{a_1,a_2,\dots\}$ and $\{b_1,b_2,\dots\}$, respectively. As we set $\alpha =0$, $\hat{T}_A$ is the average backoff time in the response phase, and $\hat{T}_B$ is that in the initiation phase.

We now introduce our mechanism for a node $a_{i^*}$. Node $a_{i^*}$ first guesses that the average amount of backoff time in the response phase is $\hat{T}_{a_{i^*},0}$, and sets $\lambda_{a_{i^*}}=\frac{1}{\hat{T}_{a_{i^*},0}}-\frac{w_{a_{i^*}}}{g_{a_{i^*}}}$. Node $a_{i^*}$ then observes the system behaviors and updates its value of $\lambda_{a_{i^*}}$ every $M$ rounds. When ${a_{i^*}}$ updates $\lambda_{a_{i^*}}$ the $k^{th}$ time, it computes the average backoff time in the response phase since it last updates its values, denoted by $\hat{T}_{A,k-1}$. Node ${a_{i^*}}$ then sets $\hat{T}_{a_{i^*},k}$ so that $$\frac{1}{\hat{T}_{a_{i^*},k}}=\frac{1}{\hat{T}_{a_{i^*},k-1}}-\delta_k(\frac{1}{\hat{T}_{A,k-1}}-\frac{1}{\hat{T}_{a_{i^*},k-1}}),$$ and $\lambda_{a_{i^*}}=\frac{1}{\hat{T}_{a_{i^*},k}}-\frac{w_{a_{i^*}}}{g_{a_{i^*}}}$, where the values of $\delta_k$ are chosen so that $\sum_{k=1}^\infty\delta_k=\infty$ and $\sum_{k=1}^\infty\delta^2_k<\infty$. For example, one can choose $\delta_k=\frac{\epsilon}{k}$, where $\epsilon$ is a small constant.  The mechanism for a node $b_{j^*}$ can be derived similarly. The only difference is that node $b_{j^*}$ updates its value of $\gamma_{b_{j^*}}$ based on the average backoff time in the initiation phase.

As described in the following theorems, this mechanism has two important features. First, this mechanism is compatible to the node's incentive. Second, this mechanism converges to a Nash Equilibrium. %We omit the proofs due to space limitation. Detailed proofs can be found in \cite{arxiv}.

%We first show that this mechanism is compatible to the node's incentive.

\begin{theorem} \label{theorem:distributed:incentiveA}
Fix the values of $\gamma_n$ and $\lambda_n$, for all $n\neq a_{i^*}$, such that $\gamma_{a_i}=0$, for all $i$, and $\lambda_{b_j}=0$, for all $j$. The average total cost of node $a_{i^*}$ is smaller when it sets $\lambda_{a_{i^*}}=\frac{1}{\hat{T}_{a_{i^*},k+1}}-\frac{w_{a_{i^*}}}{g_{a_{i^*}}}$, than when it sets $\lambda_{a_{i^*}}=\frac{1}{\hat{T}_{a_{i^*},k}}-\frac{w_{a_{i^*}}}{g_{a_{i^*}}}$, for all $k$.
\end{theorem}
\begin{IEEEproof}
Since we impose $\gamma_{a_{i^*}}=0$, node $a_{i^*}$ has no control on the amount of backoff time in the initiation phase. Hence, it suffices to show that setting $\lambda_{a_{i^*}}=\frac{1}{\hat{T}_{a_{i^*},k+1}}-\frac{w_{a_{i^*}}}{g_{a_{i^*}}}$ achieves smaller average additional total cost in the response phase than setting $\lambda_{a_{i^*}}=\frac{1}{\hat{T}_{a_{i^*},k}}-\frac{w_{a_{i^*}}}{g_{a_{i^*}}}$.

We have $\hat{T}_{A,k}=\frac{1}{\sum_{i\neq i^*}\lambda_{a_{i^*}}+(\frac{1}{\hat{T}_{a_{i^*},k}}-\frac{w_{a_{i^*}}}{g_{a_{i^*}}})}$, and $g_{a_{i^*}}-\frac{w_{a_{i^*}}}{\sum_{i\neq i^*}\lambda_{a_{i^*}}}=\frac{g_{a_{i^*}}}{\sum_{i\neq i^*}\lambda_{a_{i^*}}}(\frac{1}{\hat{T}_{A,k}}-\frac{1}{\hat{T}_{a_{i^*},k}})$. By (\ref{equation:bilateral:costA}), if $g_{a_{i^*}}-\frac{w_{a_{i^*}}}{\sum_{i\neq i^*}\lambda_{a_{i^*}}}>0$, the expected additional cost of $a_{i^*}$ strictly decreases with the backoff timer of $a_{i^*}$, $t$. Moreover, if $g_{a_{i^*}}-\frac{w_{a_{i^*}}}{\sum_{i\neq i^*}\lambda_{a_{i^*}}}>0$, we have $\hat{T}_{a_{i^*},k}>\hat{T}_{A,k}$ and $\hat{T}_{a_{i^*},k}<\hat{T}_{a_{i^*},k+1}$. For every positive constant $C$, $Prob(t<C)$ is smaller when $\lambda_{a_{i^*}}=\frac{1}{\hat{T}_{a_{i^*},k+1}}-\frac{w_{a_{i^*}}}{g_{a_{i^*}}}$ than when $\lambda_{a_{i^*}}=\frac{1}{\hat{T}_{a_{i^*},k}}-\frac{w_{a_{i^*}}}{g_{a_{i^*}}}$. Therefore, the average additional total cost is smaller when $\lambda_{a_{i^*}}=\frac{1}{\hat{T}_{a_{i^*},k+1}}-\frac{w_{a_{i^*}}}{g_{a_{i^*}}}$ than when $\lambda_{a_{i^*}}=\frac{1}{\hat{T}_{a_{i^*},k}}-\frac{w_{a_{i^*}}}{g_{a_{i^*}}}$.

On the other hand, if $g_{a_{i^*}}-\frac{w_{a_{i^*}}}{\sum_{i\neq i^*}\lambda_{a_{i^*}}}<0$, the expected additional cost of $a_{i^*}$ strictly increases with the backoff timer of $a_{i^*}$, $t$. Moreover, if $g_{a_{i^*}}-\frac{w_{a_{i^*}}}{\sum_{i\neq i^*}\lambda_{a_{i^*}}}>0$, we have $\hat{T}_{a_{i^*},k}<\hat{T}_{a_{i^*},k+1}$, and, for every positive constant $C$, $Prob(t<C)$ is larger when $\lambda_{a_{i^*}}=\frac{1}{\hat{T}_{a_{i^*},k+1}}-\frac{w_{a_{i^*}}}{g_{a_{i^*}}}$ than when $\lambda_{a_{i^*}}=\frac{1}{\hat{T}_{a_{i^*},k}}-\frac{w_{a_{i^*}}}{g_{a_{i^*}}}$. Therefore, the average additional total cost is also smaller when $\lambda_{a_{i^*}}=\frac{1}{\hat{T}_{a_{i^*},k+1}}-\frac{w_{a_{i^*}}}{g_{a_{i^*}}}$ than when $\lambda_{a_{i^*}}=\frac{1}{\hat{T}_{a_{i^*},k}}-\frac{w_{a_{i^*}}}{g_{a_{i^*}}}$.
\end{IEEEproof}

\begin{theorem} \label{theorem:distributed:incentiveB}
Fix the values of $\gamma_n$ and $\lambda_n$, for all $n\neq b_{j^*}$, such that $\gamma_{a_i}=0$, for all $i$, and $\lambda_{b_j}=0$, for all $j$. The average total cost of node $b_{j^*}$ becomes smaller when it updates its $\gamma_{b_{j^*}}$.
\end{theorem}
\begin{IEEEproof}
The proof is very similar to that of Theorem \ref{theorem:distributed:incentiveA}, and is hence omitted.
\end{IEEEproof}

Next, we show that our mechanism converges to the Nash Equilibrium.

\begin{theorem} \label{theorem:distributed:convergence}
If all nodes apply the proposed mechanism, then the value of $\hat{T}_{A,k}$ and $\hat{T}_{B,k}$ after each node updates $k$ times converge to $\hat{T}_A=\frac{I-1}{\sum_i\frac{w_{a_i}}{g_{a_i}}}$ and $\hat{T}_B=\frac{J-1}{\sum_j\frac{w_{b_j}}{g_{b_j}}}$, respectively, as $k\rightarrow\infty$.
\end{theorem}
\begin{IEEEproof}
We only prove that $\hat{T}_{A,k}$ converges to $\hat{T}_{A}$. Under our mechanism, the value of $\lambda_{a_i}$ after node $a_i$ updates $k$ times is
\begin{align*}
\lambda_{a_i}=&\frac{1}{\hat{T}_{a_i,k}}-\frac{w_{a_i}}{g_{a_i}}\\
=&(1+\delta_k)\frac{1}{\hat{T}_{a_i,k-1}}-\delta_k\frac{1}{\hat{T}_{A,k-1}}-\frac{w_{a_i}}{g_{a_i}}\\
=&(1+\delta_k)\lambda_{a_i,k-1}-\delta_k(\frac{1}{\hat{T}_{A,k-1}}-\frac{w_{a_i}}{g_{a_i}}),
\end{align*}
and hence
\begin{align*}
\frac{1}{\hat{T}_{A,k}}&=\sum_i\lambda_{a_i}\\
&=(1+\delta_k)(\sum_i\lambda_{a_i,k-1})-\delta_k(\frac{I}{\hat{T}_{A,k-1}}-\sum_i\frac{w_{a_i}}{g_{a_i}})\\
&=\frac{1}{\hat{T}_{A,k-1}}+\delta_k(I-1)(\frac{1}{\hat{T}_A}-\frac{1}{\hat{T}_{A,k-1}}).
\end{align*}

As we have $\sum_k\delta_k=\infty$ and $\sum_k\delta_k^2<\infty$, $\frac{1}{\hat{T}_{A,k}}$ converges to $\frac{1}{\hat{T}_A}$.
\end{IEEEproof}

\section{Nash Equilibriums for Multiple File Exchanges with Network Coding} \label{section:network coding}

In this section, we derive the Nash Equilibriums for systems that incorporate network coding and have more than two files. We use $I_A, I_B,I_C,\dots$ to denote the number of nodes that need $A,B,C,\dots$, respectively. We will show that there exist a Nash Equilibrium  where each node $n$ chooses each of its backoff timers in both phases as an exponential random variable. A lot of derivations in this section are similar to those in Section \ref{section:bilateral}. Hence, we omit some details and only report the major results in this section due to space limit.

We first consider the nodes' strategies in the response phase. Assume that a node $n$ with $X_n=X$ has sent the control packet in the initiation phase. Each node $m$ with $X_m\neq X$ secretly and randomly chooses a backoff timer $\hat{t}_m\sim EXP(\lambda_{m|X})$. Note that the value of $\lambda_{m|X}$ may depend on $X$. Let $m_0$ be the node that chooses the smallest value of $\hat{t}_m$. Then, $m_0$ will transmit a data packet of $X$ after $\hat{t}_{m_0}$ time. After $m_0$ transmits the data packet, $n$ will transmit a coded packet that contains one packet from each of the files except $X$, and every node $m$ with $X_m\neq X$ can decode one packet from the transmission from $n$. Similar to the derivations of (\ref{equation:bilateral:lambdaA}), we can show that, at a Nash Equilibrium, we have
\begin{equation}
\lambda_{m|X}=\frac{1}{\sum_{Y:Y\neq X}I_Y-1}\sum_{l:X_l\neq X}\frac{w_{l}}{g_l}-\frac{w_m}{g_m},
\end{equation}
for all $m$ such that $X_m\neq X$.

Next, we consider the nodes' strategies in the initiation phase. Assume that each node $n$ chooses a backoff timer $t_n\sim EXP(\gamma_n)$. Let $\Gamma_X:=\sum_{n:X_n=X}\gamma_n$. Similar to the derivations of (\ref{equation:bilateral:eqA}), (\ref{equation:bilateral:eqB}), and (\ref{equation:bilateral:sum}), we have that, at a Nash Equilibrium:
\begin{equation}\label{equation:network coding:gamma}
\frac{w_n}{g_n}=\frac{\Gamma_{X_n}-\gamma_n}{1-(\sum_{X:X\neq X_n}\Gamma_X)\frac{\sum_{X:X\neq X_n}I_X-1}{\sum_{m:X_m\neq X_n}{w_m}/{g_m}}},
\end{equation}
for all $n$, and
\begin{equation}\label{equation:network coding:sum}
\frac{I_X-1}{\sum_{n:X_n=X}w_n/g_n}\Gamma_X+\frac{\sum_{Y:Y\neq X}I_Y-1}{\sum_{m:X_m\neq X}{w_m}/{g_m}}\sum_{Y:Y\neq X}\Gamma_Y=1,
\end{equation}
for all $X$. (\ref{equation:network coding:sum}) represents a series of linear equations where both the number of unknowns, $\{\Gamma_X\}$, and the number of equations equal to the number of files. We can use standard techniques for solving linear equations to obtain a solution of $\{\Gamma_X\}$ to (\ref{equation:network coding:sum}). We can then use $\{\Gamma_X\}$ to obtain the values of $\{\gamma_n\}$ through (\ref{equation:network coding:gamma}). The derived $\{\gamma_n\}$ forms a Nash Equilibrium.

\section{Numerical Results} \label{section:numerical}

We now present our simulation results. We first consider a system with two files, $A$ and $B$, and 20 nodes. We assume that there are 10 nodes that possess file $A$ and need file $B$, and the other 10 nodes possess $B$ and need $A$. We set $g_n=1$ for all $n$, and $w_n$ is uniformly distributed in $[1,2]$. Each node applies the distributed mechanism introduced in Section \ref{section:implementation} to update its strategy. Each node $n$ sets $\hat{T}_{n,0}=\frac{99}{100(w_n/g_n)}$, that is, it guesses that there are 100 nodes in its group, an overestimate by a factor of 10, and all nodes in its group have the same values of $w_n$ and $g_n$ as itself. Also, each node sets $\delta_k=0.1/k$.

\begin{figure}[t]
\includegraphics[width = 3.2in]{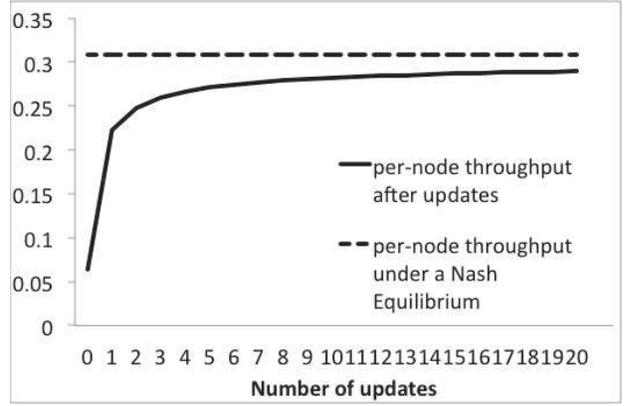}
\caption{Per-node throughput with two files.} \label{fig:numerical:converge}
\end{figure}

Fig. \ref{fig:numerical:converge} shows the resulting per-node throughput after each update. It can be shown that, even though the initial strategies of nodes are far from the Nash Equilibrium, the per-node throughput under our mechanism converges to the Nash Equilibrium very quickly. With just three updates, the per-node throughput is about 85\% of that under the Nash Equilibrium.

%Next, we consider systems with two files and variable number of nodes. For each system, half of the nodes possess $A$ and need $B$, while the other half of nodes possess $B$ and need $A$. We set $g_n=1$ for all $n$ and $w_n$ uniformly distributed in $[1,2]$. When setting the initial value $\hat{T}_{n,0}$, each node overestimates the number of nodes in its group by a factor of 10, and assumes all nodes have the same values of $g_n$ and $w_n$ as itself.

%Fig. \ref{fig:numerical:size} shows the system throughput after each node updates its strategy 10 times. Although the number of updates is small, the system throughput is very close to that under the Nash Equilibrium, especially when the number of nodes is large. These results suggest that the distributed mechanism in Section \ref{section:implementation} converges very quickly. It can also be shown that the resulting system throughput is about 0.6 packet per unit time. As the maximum possible system throughput is 1 packet per unit time, the price of anarchy on system throughput is about 1.5.

%\begin{figure}[t]
%\includegraphics[width = 3.2in]{size.eps}
%\caption{System throughput of variable number of nodes.} \label{fig:numerical:size}
%\end{figure}

Next, we consider systems that have more than two files and employ network coding. We assume that, for each file, there are ten nodes that need it. Therefore, there are a total number of $10\times\{\mbox{number of files}\}$ nodes in the system. We also set $g_n=1$ for all nodes, and $w_n$ uniformly distributed in $[1,2]$. We use the procedure described in Section \ref{section:network coding} to derive the values of $\gamma_n$ and $\lambda_n$ for all $n$, based on which we calculate the per-node throughput under a Nash Equilibrium.

\begin{figure}[t]
\includegraphics[width = 3.2in]{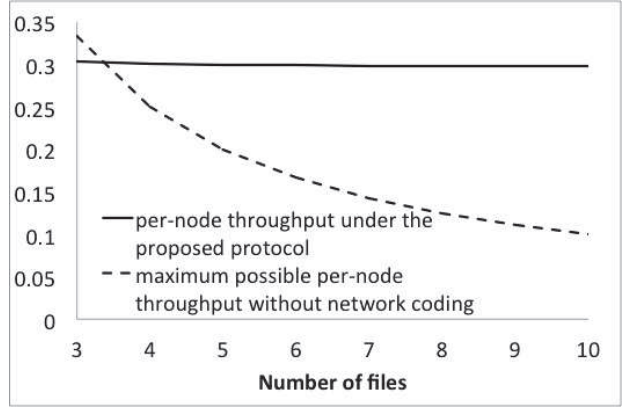}
\caption{Per-node throughput with multiple files and network coding.} \label{fig:numerical:coding}
\end{figure}

Fig. \ref{fig:numerical:coding} shows the per-node throughput under various numbers of files in the system. We compare the performance of our protocol against the maximum possible per-node throughput when network coding is not employed and when all nodes are cooperative. Without network coding, each transmission only contains one packet of one file. Hence, the maximum possible per-node throughput is $1/\{\mbox{number of files}\}$. Fig. \ref{fig:numerical:coding} shows that, when there are only three files in the system, our protocol has slightly worse per-node throughput than the case when network coding is not employed and all nodes are cooperative. This is because our protocol considers the selfish behaviors of nodes and the times spent on the two phases of backoffs, while the compared scenario assumes all nodes are cooperative and hence there is no time spent on backoffs. However, as the number of files increases, the benefits of network coding outweigh the prices of anarchy. As a result, our protocol achieves better per-node throughput than the scenario where network coding is not employed and all nodes are cooperative.

\section{Conclusions}   \label{section:conclusion}

We have discussed the problem of content distribution over wireless P2P networks. We have proposed a model that considers both the broadcast nature of wireless transmissions and the incentives of nodes. Based on the mode, a protocol for content distribution has been designed. The protocol encourages nodes to contribute to the network and is non-monetary. We have studied the performance of our protocol when all nodes are selfish. For systems with only two files, we have demonstrated that there are closed-form expressions for Nash Equilibriums and prices of anarchy. We have also proposed a distributed mechanism where all nodes update their strategies only based on their respective private information and the history of the system. We have provided numerical results that show that this mechanism converges to Nash Equilibriums very quickly. For systems with more than two files, we propose a simple extension of our protocol to incorporate network coding. We then describe a procedure to compute each node's strategy under a Nash Equilibrium. Numerical results show that our protocol may achieve better performance than scenarios where nodes are cooperative but do not employ network coding.

%For systems with more than two files, we have provided studies that suggest that the prices of anarchy are likely to be much smaller than those for systems with only two files.

\def\baselinestretch{0.85}
\small
\bibliographystyle{ieeetr}
\bibliography{reference}

\end{document}